\documentclass{ieeeaccess}
\usepackage{cite}
\usepackage{amsmath,amssymb,amsfonts}
\usepackage[binary-units=true]{siunitx}
\usepackage{url}
\usepackage{algorithmic}
\usepackage{graphicx}
\usepackage{textcomp}
\usepackage{arydshln}

\usepackage{xfp}
\NewDocumentCommand{\mean}{o m}{$
  \fpeval{round(({#2})/5,1)}
  $}
  \NewDocumentCommand{\sumL}{o m}{$
  \fpeval{round(({#2}),1)}
  $}
  \NewDocumentCommand{\sumbf}{o m}{$ \mathbf{
  \fpeval{round(({#2}),1)}}
  $}

\usepackage{multirow}

\newcommand{\CNNtime}{CNN$_{\text{time}}$}
\newcommand{\CNNfft}{CNN$_{\text{fft}}^\dagger$}
\newcommand{\FcNNfft}{FcNN$_{\text{fft}}$}
\newcommand{\FcNNpca}{FcNN$_{\text{pca}}$}

\def\BibTeX{{\rm B\kern-.05em{\sc i\kern-.025em b}\kern-.08em
    T\kern-.1667em\lower.7ex\hbox{E}\kern-.125emX}}
\begin{document}
\history{This paper is a corrected version of~\cite{seppi2021deep}. A brief overview of the issues and corrections can be found in~\cite{seppi2024corrections}.}
\doi{}

\title{Deep-Learning Approach for Tissue Classification using Acoustic Waves during Ablation with an Er:YAG Laser (Updated)}

\author{Carlo~Seppi, 
Philippe~C.~Cattin}

\address{Center for medical Image Analysis \& Navigation (CIAN), Department of Biomedical Engineering, University of Basel, 4123 Allschwil, Switzerland}

\tfootnote{We acknowledge the funding of the Werner Siemens Foundation through the \mbox{MIRACLE} (Minimally Invasive Robot-Assisted Computer-guided LaserosteotomE) project.}

\corresp{Corresponding author: carlo.seppi@unibas.ch}

\begin{abstract}
Today's mechanical tools for bone cutting (osteotomy) cause mechanical trauma that prolongs the healing process.
Medical device manufacturers constantly strive to improve their tools to further minimize the trauma.
One example of such a new tool and procedure is minimally invasive surgery using a laser as the cutting element. 
This setup allows tissue to be ablated using laser light instead of mechanical tools, which reduces the healing time after surgery. 

A reliable feedback system is crucial during laser surgery to avoid collateral damage to surrounding tissues.
Therefore, we propose a tissue classification method that analyzes the acoustic waves generated during laser ablation and demonstrates its applicability in an ex-vivo experiment.

The ablation process with a microsecond pulsed Er:YAG laser produces acoustic waves, which were acquired with an air-coupled transducer.
We then used these acquired waves to classify five porcine tissue types: hard bone, soft bone, muscle, fat, and skin.
For automated tissue classification of the measured acoustic waves, we compared five Neural Network~(NN) approaches: a one-dimensional Convolutional Neural Network~(CNN) with a time-dependent input, a Fully-connected Neural Network~(FcNN) with either the frequency spectrum or the principal components of the frequency spectrum as an input, and a combination of a CNN and an FcNN with the time-dependent data and its frequency spectrum as an input. 
In addition, several consecutive acoustic waves were used to improve the classification task. 
We used Grad-Cam to find the activation map of the frequencies and concluded that the low-frequencies were the most important ones for this classification task.
Our results indicated that the highest accuracy of the classification task~(65.5\%-75.5\%) could be achieved by combining the time-dependent data with its frequency spectrum. 
In addition, we showed that it was sufficient to use the frequency spectrum as input and that no additional benefit was gained by applying Principal Components Analysis~(PCA) to the frequency spectrum.

\subsubsection*{Remark:} This paper is an updated version of our previous work titled \textit{Deep-Learning Approach for Tissue Classification using Acoustic Waves during Ablation with an Er:YAG Laser} originally published in~\cite{seppi2021deep}. This update addresses several issues and incorporates corrections as outlined in~\cite{seppi2024corrections}. We provide here a detailed description of our experiments and the new models we used. Therefore, some of the figures and text passages are similar or the same as in the original publication. We also provide the GitHub link to the code \footnote{\url{https://github.com/fridolinvii/tissue-classification-using-acoustic-waves-during-ablation-with-an-eryag-laser}}.

\end{abstract}

\begin{keywords}
Acoustic Feedback, Laser Ablation, Tissue Classification, Neural Network
\end{keywords}

\titlepgskip=-15pt

\maketitle

\section{Introduction}
\label{sec:introduction}
\PARstart{M}{Minimally} invasive procedures demonstrate a significant step toward accelerated recovery after surgery~\cite{hu2009comparative,luketich2003minimally}: replacing the mechanical tools in open osteotomies with laser-based ablation~\cite{lo2012femtosecond} shows a further reduction in recovery time~\cite{Baek2020, Augello2018}.
Mechanical tools -- which are still the standard in conventional osteotomy -- induce thermal and mechanical trauma due to mechanical friction.
Replacing mechanical tools with lasers can reduce this trauma~\cite{lo2012femtosecond,visuri1996erbium,baek2015comparative}.

When tissue is exposed to a microsecond pulsed Erbium-doped Yttrium Aluminium Garnet~(Er:YAG) laser, the water in the tissue is heated until it evaporates.
This process takes place in microseconds and builds up pressure that is released in a series of micro-explosions.
The explosions ablate a small portion of the tissue~\cite{kang2007hard}, creating an acoustic wave~\cite{kenhagho2018comparison}.
A transducer can then measure the resulting acoustic wave.
As carbonization causes thermal damage and reduces cutting efficiency, the ablated tissue must be continuously rehydrated and cooled ~\cite{lo2012femtosecond,visuri1996erbium,baek2015comparative,bernal2018performances,nguendon2017characterization,abbasi2017effect}.

The goal of the flagship project 
MIRACLE\footnote{MIRACLE\,\,(Minimally\,\,Invasive\,\,\mbox{Robot-Assisted}\,\,\mbox{Computer-guided} LaserosteotomE),\,\,\textit{01.08.2020},\,\,\url{ https://dbe.unibas.ch/en/research/flagship-project-miracle}}
is to improve laser osteotomy by integrating the advantages of robot-assisted laser surgery into an endoscope~\cite{rauter2020miracle,eugster2017positioning,eugster2018parallel}.
This allows the surgeon to perform laser-assisted osteotomies by inserting an endoscope into the body through small incisions or natural orifices.
Information about the environment around the endoscope, such as the type of tissue being ablated, can help the surgeon avoid cutting the wrong tissue.
Several approaches have been considered to distinguish porcine tissue types as feedback for laser tissue ablation, e.g., using optical spectroscopy~\cite{abbasi2018differentiation,zam2011optical,stelzle2011optical,abbasi2020combined} or optical coherence tomography~(OCT)~\cite{bayhaqi2019neural,bayhaqi2019fast,bayhaqi2021deep,bayhaqi2023real}.
The authors of~\cite{grad1994optoacoustic} proposed optoacoustic imaging to differentiate between different types of hard dental tissue.
In~\cite{tangermann2003sensor,rupprecht2004sensor}, the authors studied optical and acoustic signals during Er:YAG laser osteotomy.
They proposed a heuristic to decide when to turn off the laser to avoid nerve damage.
In contrast, we aim to use only the acoustic signal for tissue classification to prevent the laser from continuing the cut when tissue that should not be damaged is detected.
Similar approaches have been proposed in~\cite{nguendon2017characterization,kenhagho2019optoacoustic,herve2019}.
Another approach~\cite{nahum2018bone,nahum2019joint} uses acoustic waves in a 2D simulation to infer the acoustic density within a region of interest.
This information can then be used to classify the underlying tissue.

For this research, we used supervised deep learning to train NNs that can infer the tissue from the acoustic waves.
In a simplified ex-vivo experiment, we prepared samples of porcine tissues so that each tissue type could be ablated without interference from others.

Neural networks~\cite{bengio2013representation} have found their way into many related applications, such as medical image classification problems~\cite{deepa2011survey,li2014medical,kumar2016ensemble} or
speech and signal processing~\cite{miller1992review,kiranyaz2015real,han2016convolutional,kamper2016deep}.
Similarly, we used NNs to classify different types of porcine tissue.
In our case, the shock wave emitted during ablation was used for our classifying NNs.
Four different variations as input were compared: time-dependent pressure variation, frequency-dependent pressure variation, the combination of both, or the principal components of the frequency spectrum.
For this purpose, we used either a one-dimensional CNN, FcNN, or a combination of both (CNN+FcNN). 
Grad-Cam was used to find activation maps to analyze the frequency domain further.
Grad-Cam~\cite{selvaraju2017grad} can compute the activation maps that highlight the essential part of the data for a specific classification task.
Therefore, Grad-Cam was applied to a CNN with frequency-dependent data and found the corresponding activation map in the frequency domain.
The proposed new approach showed superior results compared to the original method~\cite{herve2019}. 
In fact, we show that using the frequency spectrum as an input to the NNs outperforms using the principal components of the frequency spectrum.
The best-performing network used the time-dependent data and its frequency spectrum as an input to the NN.
The different networks were optimized using a hyperparameter search~\cite{optuna_2019}, and the best-performing networks were then used for comparison and further analysis.


\section{Material}
In this section, we describe the setup and data collection of our experiments.

\subsection*{Setup}
\begin{figure}[t]
    \centering
    \includegraphics[width=.475\textwidth]{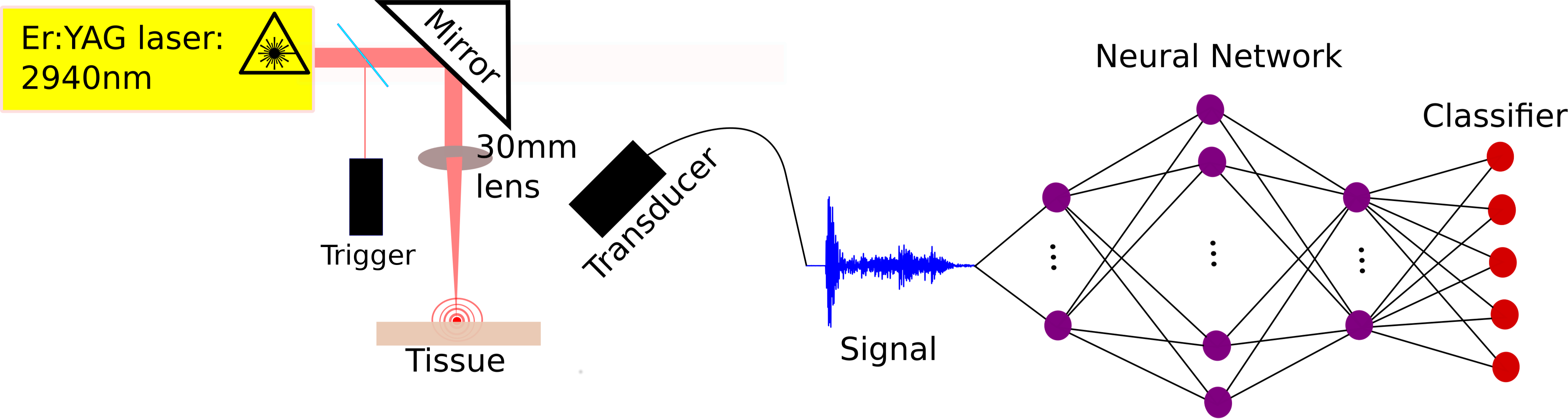}
    \caption{Setup contained an Er:YAG laser with a wavelength of $\SI{2940}{\nano\meter}$,
    where the pulses had a repetition rate of $\SI{2}{\hertz}$, and a duty cycle of \SI{260}{\micro\second} 
    The tissue was placed at the focal point of the laser, at a distance of $\SI{30}{\milli\meter}$ from the lens.
  A transducer measured the acoustic wave generated during the ablation at a distance of $\SI{5}{\centi\meter}$, with an angle of $\SI{45}{\degree}$.
    We used the measured signal as the input to our NNs to classify different tissue types.}
    \label{fig:laser_setup}
\end{figure}

Figure~\ref{fig:laser_setup} visualizes the setup used in this research.
We used an Er:YAG laser~(Syneron Candela, litetouch LI-FG0001A)
with a wavelength of $\SI{2940}{\nano\meter}$, and a pulse duration of $\SI{300}{\micro\second}$.
A $CaF_2$ mirror was placed at a small angle in front of the laser head to split the laser light into two parts.
The laser light ablated the tissue, while a small percentage of the light was reflected. 
A fast PbSe photodiode~(PbSe Fixed Gain Detector, PDA20H, $1500-\SI{4800}{\nano\meter}$) detected the reflected light that was used as the trigger to activate the measurement of the acoustic signal received by the transducer. 
The custom-built air-coupled piezoelectric transducer\footnote{provided by Tomas E. Gomez Alvarez-Arenas at the ITEFI-Instituto de Tecnologias Fisicas y de la Information, CSIC, Madrid, Spain} 
with a diameter of $\SI{15}{\milli\meter}$, a frequency range of $0.1-\SI{0.8}{\mega\hertz}$, and a resonance frequency at $\SI{0.4}{\mega\hertz}$ captured the acoustic waves generated during ablation and recorded a time window of $\SI{2.3}{\milli\second}$.
The experiment was carried out in wet conditions, utilizing a manually operated spray of distilled water on the ablation area. This approach was adopted to reduce carbonization during the laser ablation process.
The transducer was placed at a distance of $\SI{5}{\cm}$ at an angle of $\SI{45}{\degree}$ to the sample and converted the measured pressure variation into a digital signal with a sampling rate of $\SI{7.8125}{\mega\hertz}$.
The measured data was then used as input to our proposed tissue classifier.

\subsection*{Data}
\begin{figure}[t]
    \centering
    \includegraphics[width=.475\textwidth]{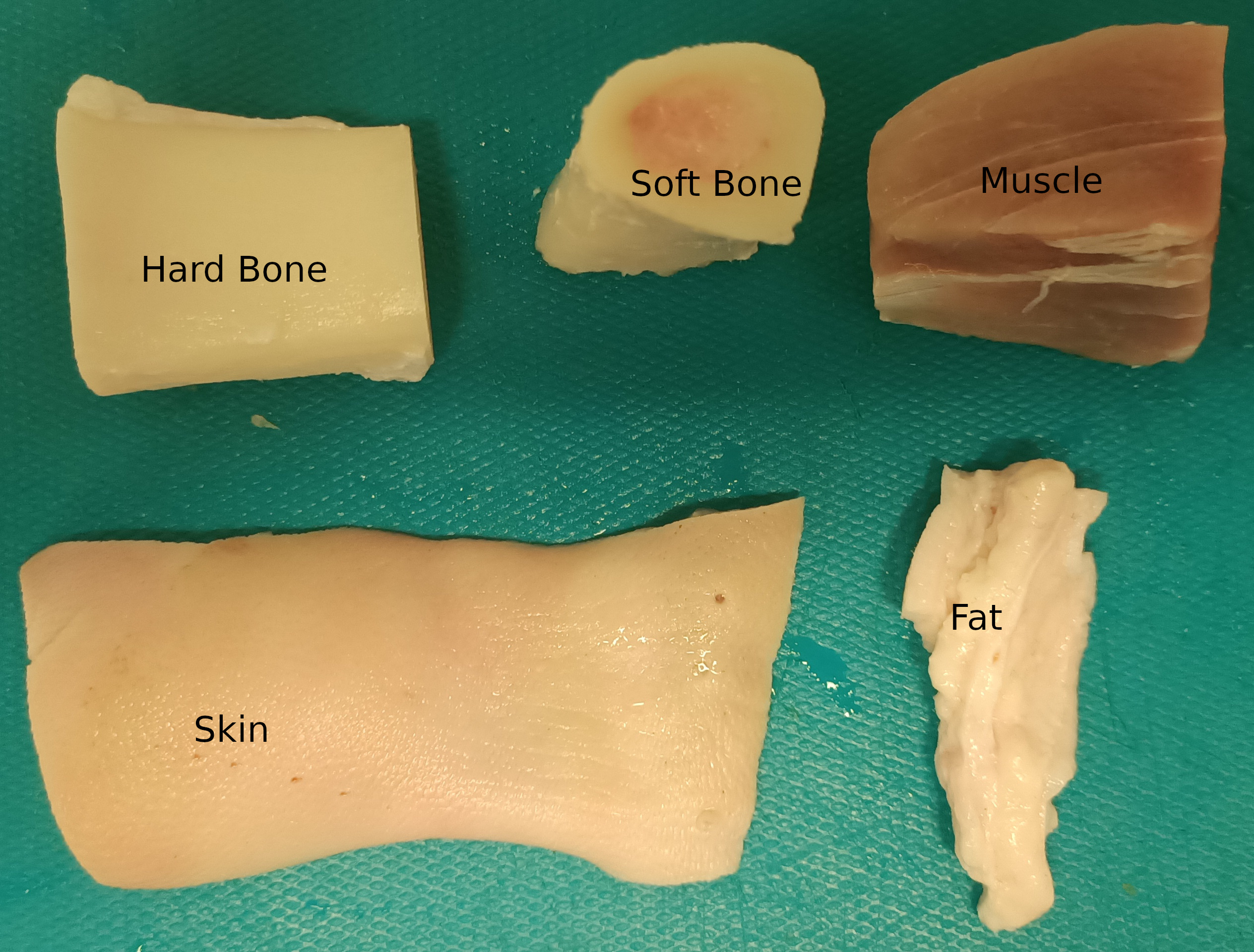}
    \caption{Five exemplary tissues used for the experiments: A total of ten samples of each tissue were used. Five 100-pulse ablations were performed on each tissue.}
    \label{fig:tissue}
\end{figure}

Data were obtained from freshly excised specimens of hard bone (compact bone fragment), soft bone (bone marrow), muscle, fat, and skin tissue.
All tissues were kept in water prior to the experiment.
The muscle was carefully removed from the knuckle of the pig. 
We used the rim area of the hard bone as the specimen and the center of the bone as the soft bone specimen. 
The skin and fat specimens were dissected from the pig foot. 
In Figure~\ref{fig:tissue}, shows some exemplary tissues used for our experiments.

\begin{figure}[t]
    \centering
    \includegraphics[width=.475\textwidth]{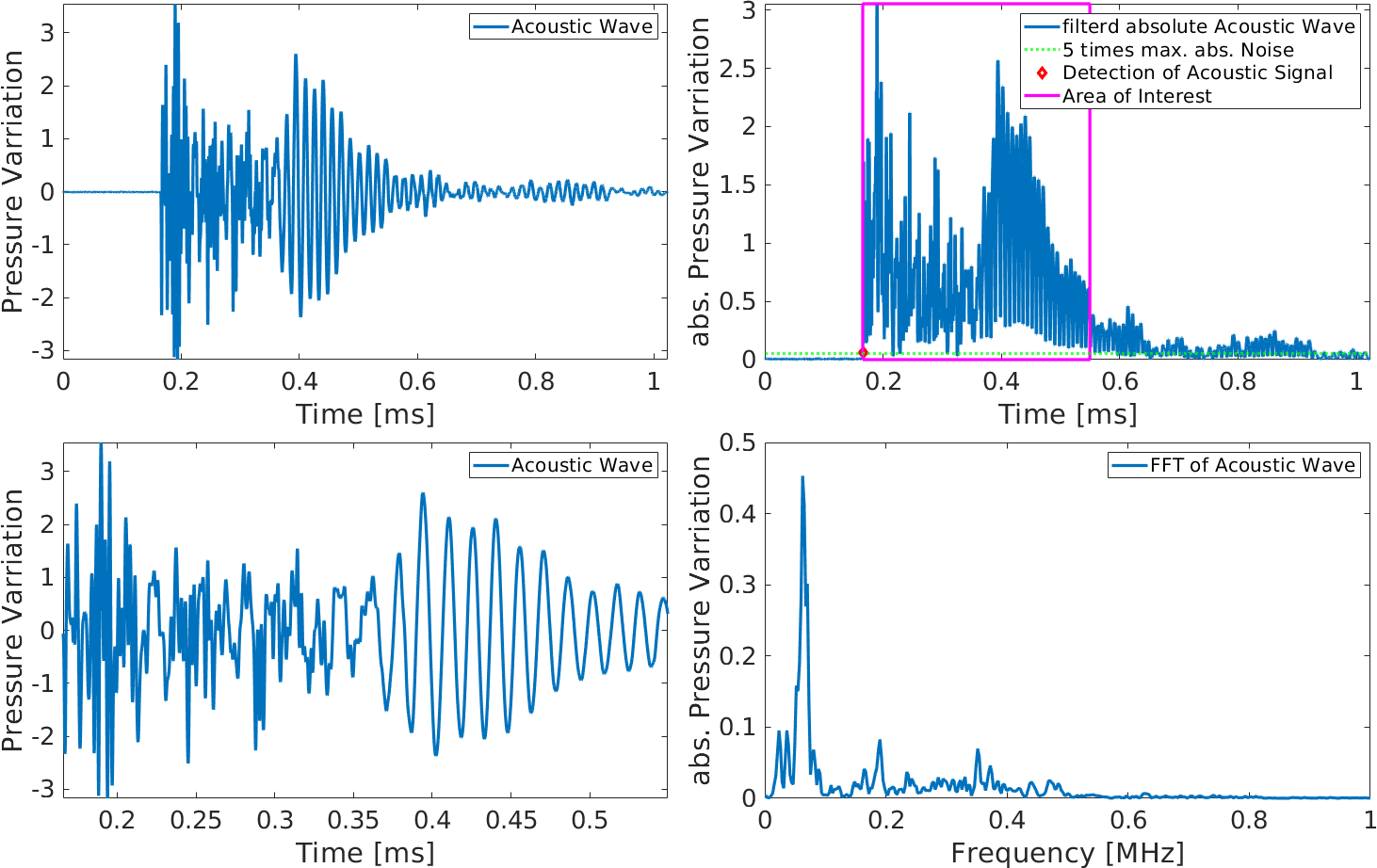}
    \caption{Visualization of the preprocessing steps: Top left, a $\SI{1}{\milli\second}$ window shows the measured acoustic wave from a soft bone ablation. The top right is the absolute median filtered acoustic wave (kernel size 11). To remove the ToF, we take five times the maximum value of the first $\SI{96}{\micro\second}$ of the filtered wave (green dotted line). The intersection is marked with a red dot, where the area of interest (marked in magenta) starts with a window size of $\SI{384}{\micro\second}$. 
    The bottom left shows the resulting acoustic wave in the area of interest, which we used as an input for our NNs. The bottom right is the frequency spectrum of $0-\SI{1}{\mega\Hz}$ of the acoustic wave shown on the bottom left. These frequency spectra are used as input for our NNs.    
    }
    \label{fig:preprocessing}
\end{figure}

Ten specimens of each tissue type were probed, each being ablated five times, creating a perpendicular hole to the tissue surface.
Each laser ablation consisted of $100$ laser pulses at a repetition rate of $\SI{2}{\hertz}$.
Consequently, $500$ measurements of ablation-induced pressure variations were made for each specimen, giving a total of $\num{5000}$ measurements per tissue type.
Since we repeated the experiment for five tissue types, it resulted in a total of $\num{25000}$ measurements. 
The data acquisition window for each acoustic wave was $\SI{2.3}{\milli\second}$. 
This paper employed a method similar to that described in~\cite{seppi2022bone} to eliminate time-of-flight~(ToF) information, using a time window of size $\SI{384}{\micro\second}$ (see Figure~\ref{fig:preprocessing}); and with a frame rate of $\SI{7.8125}{\mega\hertz}$ each acoustic wave was therefore represented as a $1\times3000$\,-dimensional array.  

\section{Methods}
We aim to interpret the acoustic waves generated during ablation to classify the ablated tissue using an end-to-end neural network approach. 
To this end, the data set was divided into three disjoint subsets: \SI{60}{\percent} (6 specimens of each tissue type) were used for training, \SI{20}{\percent} (2 specimens of each tissue type) for validation, and the remaining \SI{20}{\percent} (2 specimens of each tissue type) as test data.  
After each training epoch, which was an iteration of all training data during the network training, the performance of the algorithm was estimated on the validation data.
To detect overfitting and ensure the robustness of our network, we used only previously unseen data for testing, i.e., measurements from a given sample were used only in one subset.
To evaluate the robustness and variability of our approach, a fivefold cross-validation was performed. 
To do this, the data was divided into five disjoint subsets,~e.g.,\,\mbox{$\mathcal{A}$, $\mathcal{B}$, $\mathcal{C}$, $\mathcal{D}$, $\mathcal{E}$}, where each subset contains two specimens of each tissue type. 
The first network used the subset \mbox{$\mathcal{A}\cup\mathcal{B}\cup\mathcal{C}$} for training, $\mathcal{D}$ for validation, and \mbox{$\mathcal{E}$} for testing.
Note that the training, validation, and test data sets were disjoint.
The second network used the subset \mbox{$\mathcal{B}\cup\mathcal{C}\cup\mathcal{D}$} for training, $\mathcal{E}$ for validation, and \mbox{$\mathcal{A}$} for testing. 
This continued in a rotating fashion until the fifth network used \mbox{$\mathcal{E}\cup\mathcal{A}\cup\mathcal{B}$} for training, $\mathcal{C}$ for validation, and \mbox{$\mathcal{D}$} for testing.
As input for the NNs, we used either 1, 5, or 10 consecutive acoustic waves.
For our proposed NNs, a hyperparameter search with 10 consecutive acoustic waves on one of the subsets with the parameters shown in Table~\ref{tab:hyperparameter} was performed.
The network was trained using the same parameters for all cross-validations, as well as for those with 1 and 5 acoustic waves as input. The optimizer used during training was Adam~\cite{kingma2014adam}.

\begin{figure}
    \centering
    \footnotesize
    \textbf{Time-dependent (\CNNtime)}
    \begin{itemize}
        \item[(0)] $N$ BatchNorm1d(3000) (one for each time-dependent wave)
        \item[(1a)] Conv1d($N$, 8, kernel\_size=(128,), stride=(1,))
        \item[(1b)] MaxPool1d(kernel\_size=8, stride=8)
        \item[(2a)] Conv1d(8,16, kernel\_size=(128,), stride=(1,))
        \item[(2b)] MaxPool1d(kernel\_size=8, stride=8)
        \item[(3)] Flattening(464)
        \item[(4)] BatchNorm1d(464)
        \item[(5a)] Dropout(p=0.3923535702072316)
        \item[(5b)] Linear(in\_features=464, out\_features=16)
        \item[(6a)] Dropout(p=0.3923535702072316)
        \item[(6b)] Linear(in\_features=16, out\_features=5)
        \item[]
    \end{itemize} 
    
    \textbf{Frequnecy-dependent (\FcNNfft)\quad (for Grad-Cam $\dagger$: \CNNfft)}
    \begin{itemize}
        \item[(0)] $N$ BatchNorm1d($M$) (one for each frequency-dependent wave)
        \item[($\dagger$)] Conv1d($N$, $N$, kernel\_size=(1,), stride=(1,))
        \item[(1)] Flattening($M\cdot N$)
        \item[(2a)] Dropout(p=0.4370396877827484)
        \item[(2b)] Linear(in\_features=$M\cdot N$, out\_features=256)
        \item[(3a)] Dropout(p=0.4370396877827484,)
        \item[(3b)] Linear(in\_features=256, out\_features=256)
        \item[(4a)] Dropout(p=0.4370396877827484,)
        \item[(4b)] Linear(in\_features=256, out\_features=256)
        \item[(5a)] Dropout(p=0.4370396877827484,)
        \item[(5b)] Linear(in\_features=256, out\_features=256)
        \item[(6a)] Dropout(p=0.4370396877827484,)
        \item[(6b)] Linear(in\_features=256, out\_features=256)
        \item[(7a)] Dropout(p=0.4370396877827484,)
        \item[(7b)] Linear(in\_features=256, out\_features=5)
        \item[]
    \end{itemize} 

        \textbf{Time/Frequency-dependent (CNN+FcNN)}
    \begin{itemize}
        \item[(0a)] $N$ BatchNorm1d(3000) (one for each time-dependent wave)
        \item[(0b)] $N$ BatchNorm1d(385) (one for each frequency-dependent wave)
        \item[(1a)] Conv1d($N$, 10, kernel\_size=(128,), stride=(1,))
        \item[(1b)] MaxPool1d(kernel\_size=32, stride=32)
        \item[(2)] BatchNorm1d(890)
        \item[(3a)] Flattening($890$) (flatten time-dependent output)
        \item[(3b)] Flattening($385\cdot N$) (flatten frequency-dependent output)
        \item[(3c)] CombineData($(890,385\cdot N)$,$890+385\cdot N$)
        \item[(4a)] Dropout(p=0.361871665168048)
        \item[(4b)] Linear(in\_features=$890+385\cdot N$, out\_features=128)
        \item[(5a)] Dropout(p=0.361871665168048)
        \item[(5b)] Linear(in\_features=128, out\_features=128)
        \item[(6a)] Dropout(p=0.361871665168048)
        \item[(6b)] Linear(in\_features=128, out\_features=128)
        \item[(7a)] Dropout(p=0.361871665168048)
        \item[(7b)] Linear(in\_features=128, out\_features=5)
        \item[] 
    \end{itemize} 
    
            \textbf{Frequnecy-dependent PCA (\FcNNpca)}
    \begin{itemize}
        \item[(0)] Flattening($323\cdot N$)
        \item[(1a)] Dropout(p=0.29176279628963564)
        \item[(1b)] Linear(in\_features=$323\cdot N$, out\_features=512)
        \item[(2a)] Dropout(p=0.29176279628963564)
        \item[(2b)] Linear(in\_features=512, out\_features=512)
        \item[(3a)] Dropout(p=0.29176279628963564)
        \item[(3b)] Linear(in\_features=512, out\_features=512)
        \item[(4a)] Dropout(p=0.29176279628963564)
        \item[(4b)] Linear(in\_features=512, out\_features=512)
        \item[(5a)] Dropout(p=0.29176279628963564)
        \item[(5b)] Linear(in\_features=512, out\_features=5)
    \end{itemize} 
    \caption{Overview of the different NNs we used. $N$ is the number of consecutive acoustic waves used as an input. $M$ is the input size of the FcNNs when using frequency-dependent data: Namely for the frequency range $0-\SI{1}{\mega\Hz}$ we have $M=385$, for the low-frequency ($0-\SI{0.333}{\mega\Hz}$) and mid-frequency ($0.333-\SI{0.666}{\mega\Hz}$) is $M=128$, and for high-frequency ($0.666-\SI{1}{\mega\Hz}$) is $M=129$. 
    }
    \label{fig:model}
\end{figure}

\subsection*{Time-dependent (\CNNtime)}
The first NN we present is a CNN. The input was the time-dependent acoustic wave. 
Similar to~\cite{seppi2022bone}, we removed the ToF from the initial measurement. 
The preprocessing is visualized in Figure~\ref{fig:preprocessing}. 
First, a median filter with kernel size 11 was applied to the absolute value of the acoustic measurement. 
Then the maximum value of the first $\SI{96}{\micro\second}$ (which is just background noise) was multiplied by five. 
As soon as the filtered wave exceeded this threshold, a time window of $\SI{384}{\micro\second}$ was used as the input for the CNN. 
The input to the CNN was either 1, 5, or 10 consecutive process waves. 
Thus, the input tensor was of size $N \times 3000$, where $N\in\{1,5,10\}$.
We visualize our CNN in Figure~\ref{fig:model}.
Our CNN started with a batch normalization of each acoustic wave separately. 
A convolutional layer was applied with $N$ channels as input and 8 channels as output, using a kernel size of 128. 
This was followed by a ReLU activation and a max pooling layer with kernel size and stride of 8.
Another convolutional layer was then applied, producing 16 output channels, using a further ReLU activation and max pooling layer with a kernel size and stride of 8.
The output was flattened, followed by batch normalization and a hidden fully connected layer of 16 neurons, followed by a ReLU activation function. 
The final layer was again a fully-connected layer with five neurons as output, each representing a tissue type to be classified. 
A dropout layer was applied to each fully connected layer.
We used a batch size of 256 and a learning rate of 0.0001.

\subsection*{Frequency-dependent (\FcNNfft) }
To transform the time-dependent data into the frequency domain, the truncated time-dependent data was first subjected to a Hamming window to reduce the leakage in the FFT~\cite{downey2016think}. 
Then, we used the absolute value of the Fast Fourier Transformation~(FFT) transformation in the frequency spectrum between $\SI{0}{\mega\Hz}$ and $\SI{1}{\mega\Hz}$, as visualized on the bottom right of Figure~\ref{fig:preprocessing}.
Note that a broader frequency spectrum than the frequency range of the transducer was used to ensure that no important frequencies were omitted during the preprocessing step. 
The resulting tensor was of the size $N \times 385$ for $N\in\{1,5,10\}$, where $N$ was the number of consecutive acoustic waves used as input. 
For our FcNN visualized in Figure~\ref{fig:model} (omitting the layer marked with $\dagger$), we first applied batch normalization to each of the inputs. 
After flattening, 5 hidden fully connected layers of 256 neurons, where a ReLU activation function followed each. 
A dropout layer was added to all fully connected layers to train the neural network. The last fully connected layer had an output of 5. A batch size of 256 and a learning rate of 0.001 were used during the training process.

\subsubsection*{Grad-Cam (\CNNfft)}\label{sec:Grad-Cam}
Grad-Cam~\cite{selvaraju2017grad} uses the gradients of the convolutional layers to generate an \textit{activation map} highlighting important regions of the input data, which can help to understand which parts of the data are important for the classification task.
To highlight the important frequency ranges, a convolutional layer with a kernel size of one was added to the \FcNNfft, as indicated with a $\dagger$ in Figure~\ref{fig:model}, and named it \CNNfft.
We computed the gradient of all correctly classified tissues of the convolutional layer. 
Since the kernel size of the layer was one, each value of the gradient from the convolutional layer represents a frequency.
To obtain an array of the same size as the input, the gradient of the convolutional layer was passed through a ReLU activation function, and then the sum was taken over the number of channels.
The higher its value within the array, the more important that frequency was for the classification task. 
The results were confirmed with the \FcNNfft ~for low-frequency ($0-\SI{0.333}{\mega\Hz}$, with $M=128$), mid-frequency ($0.333-\SI{0.666}{\mega\Hz}$, with $M=128$), and high-frequency ($0.666-\SI{1}{\mega\Hz}$, with $M=129$) spectra.

\subsection*{Time/Frequency-dependent (CNN+FcNN)}
The performance improved when we combined the architectures of the time-dependent and frequency-dependent NN.
As shown in Figure~\ref{fig:model}, we applied batch normalization to each time- and frequency-dependent acoustic wave. 
A convolutional layer with a kernel size of 128 and an output channel of 10 was applied to the time-dependent input. 
The ReLU activation function was then applied, followed by a max pooling layer with a kernel and stride size of 32.
The output was flattened, and batch normalization was applied.
This was then concatenated with the flattened frequency-dependent data, followed by three hidden fully connected layers of 128 neurons, with a ReLU activation following each hidden layer. 
Again, a dropout layer was applied to each fully connected layer. 
We used a batch size of 128 and a learning rate of 0.0001. 

\subsection*{Artificial Neural Network~(ANN)}
The average of two successive time-dependent acoustic waves was first computed to compare the method with the top-performing network from \cite{herve2019}. Following this, the FFT was applied, and the first three principal components were extracted from its absolute value in the frequency range \(0-\SI{1}{\mega\Hz}\). These components served as input to a shallow neural network with a 10-neuron hidden layer. A ReLU activation function was utilized, and the final layer consisted of five neurons for tissue classification.

\subsection*{Principal Component Analysis~(\FcNNpca)}
Inspired by~\cite{herve2019}, we applied principal component analysis to the frequency-dependent data. 
We then used the 323 first principal components from the transformation as input to our NN.
As shown in Figure~\ref{fig:model}, we chose 4 hidden layers with 512 neurons, a batch size of 32, and a learning rate of 0.0001.
Again, a dropout layer was applied to each fully connected layer, and the output of the final layer was 5, which corresponds to the number of classes.

\begin{table}[t]
    \caption{We used a hyperparameter search to find a good-performing network. The table shows the parameters we tested our networks with (\FcNNfft, \FcNNpca, \CNNtime, and CNN+FcNN). Note that FcNN uses those marked with a $^*$, and that the number of principal components was also used as a hyperparameter~(pca) when it was used as an input.}
    \centering
    \begin{tabular}{l r}
         learning rate$^*$      & $\{0.1,0.01,0.001,0.0001,0.00001\}$ \\
         batch size$^*$     &  $\{16,32,64,128,256,512,1024\}$\\
         \# of FC layers$^*$ &  $\{1,2,3,4,5\}$\\
         \# of neurons$^*$   &  $\{16,32,64,128,256,512,1024\}$\\
         dropout$^*$        &   $[0,0.5]$\\
         \hdashline
         \# of conv. layers & $\{1,2,3,4,5\}$ \\
         conv. kernel size  & $\{2,4,8,16,32,64,128,256\}$ \\
         output channels    & $\{1,2,3,4,5,6,7,8,9,10\}$ \\
         max pool kernel size &$\{2,4,8,16,32,64,128\} $\\
         \hdashline
         \# of principal components~(pca) & $\{3,\hdots,380\}$
    \end{tabular}
    \label{tab:hyperparameter}
\end{table}


\section{Results}
\label{sec:results}
We implemented the networks in \textit{Python~(3.6.9)} using \textit{PyTorch~(1.10.1)}~\cite{paszke2017automatic}.
The hyperparameter search was done with \textit{optuna~(2.10.1)}~\cite{optuna_2019}.
The networks were trained until the accuracy of the validation data did not improve for ten epochs. 
Note that an epoch is an iteration over all training data. 
Additionally, fivefold cross-validation was applied, permuting the training, validation, and test data as outlined in the previous section. In the final step, previously unseen measurements (20\%, two samples for each tissue type) were used to test the network.
This allowed us to assess the robustness and generalization capabilities of the network on unseen data.
An overview of the average accuracy of the NNs is given in Table~\ref{tab:results}.

\begin{table}[t]
{\footnotesize
    \centering
    \caption{Performance comparison of the different NNs for each tissue with 1, 5, or 10 consecutive acoustic waves. The best-performing network is the $^{10}$CNN+FcNN with 10 consecutive waves. In addition, we confirm that the most important features were in the low-frequency spectrum of the acoustic wave with $^{5,L}$\FcNNfft with 5 consecutive waves. We show the mean accuracy of Hard Bone~(HB), Soft Bone~(SB), Fat~(F), Skin~(S), and Muscle~(M) of the fivefold cross-validation.}
    \begin{tabular}{r  |c c c c c | c}
     NN / [\%]          & HB & SB & F & S & M & $\varnothing$ \\
     \hline
     $^1$ANN &   
     \mean{0+       17.88+  10.51+  0+      0.00}&
     \mean{59.39+   37.98+  13.43+  45.35+  77.38}&
     \mean{77.88+   56.97+  79.09+  86.66+  59.60}&
     \mean{2.73+    12.22+  12.22+  0.0+    7.78}&
     \mean{0.00+    0+      11.01+  0.0+    0.0}&
     \mean{28.00+   25.01+  25.25+  26.40+  28.95}\\
     $^5$ANN  & 
     \mean{1.67+    36.74+  15.26+  6.53+  29.47}&
     \mean{57.67+   34.53+  14.42+  27.05+  31.58}&
     \mean{24.11+   31.26+  12.53+  9.58+   72.53}&
     \mean{11.68+   28.21+  28.00+  42.53+  24}&
     \mean{27.68+   10.95+  81.05+  70.00+  0}&
     \mean{24.44+   28.34+  30.25+  31.14+  31.52}\\
     $^{10}$ANN & 
     \mean{0.44+    27.78+  19.11+  4.67+   28.67}&
     \mean{49.89+   35.56+  26.89+  0.78+   32.44}&
     \mean{68.22+   37.89+  5.44+   3.44+   62.78}&
     \mean{14.89+   39.44+  51.00+  70.67+  28.11}&
     \mean{4.89+    7.00+   65.22+  75.67+  0}&
     \mean{27.67+   29.53+  33.53+  31.04+  30.40}\\
     \hdashline
     $^{1}$\FcNNpca &  
     \mean{45.60+   71.10+  54.10+  60.50+  47.50}&
     \mean{19.70+   51.40+  50.50+  35.40+  82.70}&
     \mean{57.50+   55.60+  61.50+  43.00+  58.60}&
     \mean{26.00+   65.50+  40.00+  54.80+  20.40}&
     \mean{63.00+   71.70+  57.40+  79.90+  64.70}&
     \mean{42.36+   63.06+  52.70+  54.72+  54.78}\\
     $^{5}$\FcNNpca & 
     \mean{68.44+   81.77+  48.02+  47.60+  63.02}&
     \mean{22.92+   35.63+  57.60+  42.92+  95.73}&
     \mean{60.31+   45.94+  77.18+  52.92+  64.06}&
     \mean{26.67+   48.23+  54.06+  72.19+  25.00}&
     \mean{63.23+   37.71+  56.77+  79.27+  55.94}&
     \mean{48.31+   49.85+  58.73+  58.98+  60.75}\\
     $^{10}$\FcNNpca &  
     \mean{66.48+   79.67+  55.16+  65.60+  63.96}&
     \mean{30.00+   64.84+  60.00+  46.48+  97.91}&
     \mean{61.54+   44.95+  80.11+  57.58+  69.12}&
     \mean{33.41+   52.20+  52.31+  62.97+  24.84}&
     \mean{55.93+   46.59+  55.05+  82.97+  40.66}&
     \mean{49.47+   57.65+  60.53+  63.12+  59.30}\\
     \hdashline
     $^{1}$\FcNNfft & 
     \mean{36.90+   70.30+  63.20+  54.10+  68.50}&
     \mean{34.00+   45.90+  62.70+  37.80+  79.80}&
     \mean{55.80+   50.90+  70.90+  53.60+  74.20}&
     \mean{18.40+   57.10+  46.90+  67.50+  28.30}&
     \mean{63.50+   71.00+  57.20+  81.80+  53.50}&
     56.1 \\ 
     $^{5}$\FcNNfft & 
     \mean{49.17+   89.27+  61.04+  59.79+  77.71}&
     \mean{38.44+   60.83+  76.25+  45.31+  89.27}&
     \mean{69.38+   57.08+  82.50+  68.75+  70.73}&
     \mean{33.54+   56.98+  58.65+  77.81+  35.42}&
     \mean{59.38+   42.29+  66.67+  86.56+  41.25}&
     \mean{49.98+   61.29+  69.02+  67.65+  62.88}\\
     $^{10}$\FcNNfft & 
     \mean{51.76+   90.88+  63.20+  54.10+  68.50}&
     \mean{44.29+   71.32+  62.70+  37.80+  79.80}&
     \mean{68.02+   60.11+  70.90+  53.60+  74.20}&
     \mean{36.04+   66.37+  46.90+  67.50+  28.30}&
     \mean{72.09+   52.09+  57.20+  81.80+  53.50}&
     \mean{54.44+   68.15+  60.18+  58.96+  60.86}\\
      \hdashline
      $^1$\CNNtime  &   
     \mean{51.10+   80.30+  58.70+  71.50+  58.80}&
     \mean{44.50+   50.30+  62.60+  60.30+  89.50}&
     \mean{57.00+   58.40+  65.00+  68.00+  70.10}&
     \mean{25.10+   73.90+  38.20+  70.50+  44.70}&
     \mean{69.40+   54.80+  67.40+  83.10+  63.90}&
     \mean{49.42+   63.54+  58.38+  70.68+  65.40} \\
     $^5$\CNNtime  &  
     \mean{66.46+   75.52+  49.79+  80.31+  68.23}&
     \mean{44.90+   76.04+  79.79+  68.75+  93.85}&
     \mean{62.40+   63.75+  71.98+  66.88+  91.25}&
     \mean{7.19+    73.85+  26.15+  76.88+  55.73}&
     \mean{83.44+   73.75+  76.98+  86.25+  64.58}&
     \mean{52.87+   72.58+  60.94+  75.81+  74.73} \\
     $^{10}$\CNNtime  & 
     \mean{76.70+   77.69+  55.60+  77.47+  62.31}&
     \mean{55.60+   86.48+  81.54+  73.19+  91.87}&
     \mean{70.11+   67.80+  80.11+  57.58+  84.95}&
     \mean{17.58+   76.15+  54.62+  78.57+  71.54}&
     \mean{79.23+   68.57+  70.33+  89.56+  65.27}&
     \mean{59.85+   75.34+  68.44+  75.27+  75.19} \\
     \hdashline
     $^1$CNN+FcNN  &   
     \mean{65.30+   83.60+  56.90+  70.20+  62.10}&
     \mean{37.90+   69.10+  62.20+  54.70+  93.60}&
     \mean{53.20+   61.50+  73.90+  74.30+  79.20}&
     \mean{22.00+   74.20+  60.40+  82.00+  56.90}&
     \mean{77.20+   70.90+  63.00+  80.80+  54.10}&
     \mean{51.12+   71.86+  63.28+  72.40+  69.18} \\
     $^5$CNN+FcNN  &  
     \mean{75.00+   85.31+  53.75+  69.38+  77.29}&
     \mean{41.15+   82.08+  76.77+  70.00+  97.40}&
     \mean{66.25+   61.15+  90.63+  77.08+  84.06}&
     \mean{32.08+   78.23+  66.66+  87.08+  54.58}&
     \mean{85.83+   72.71+  67.19+  86.35+  53.75}&
     \mean{60.06+   75.90+  71.00+  77.98+  73.42} \\
     $\mathbf{^{10}}$\textbf{CNN+FcNN}  & 
     \textbf{79.3} & \textbf{82.5} & \textbf{77.8} & \textbf{66.9} & \textbf{71} & \textbf{75.5} \\
    \hline
    \hline
    $^{5}$\CNNfft &
     \mean{53.44+   89.48+  12.81+  48.13+  71.77}&
     \mean{38.33+   48.96+  20.52+  55.52+  96.46}&
     \mean{66.98+   58.33+  57.71+  62.29+  69.38}&
     \mean{27.71+   66.15+  36.88+  72.50+  40.73}&
     \mean{78.02+   53.54+  60.21+  74.90+  41.56}&
     \mean{52.90+   63.29+  37.63+  62.67+  63.98}\\
     $\mathbf{^{5,\text{\textbf{{L}}}}}$\textbf{FcNN}$\mathbf{_{\text{\textbf{fft}}}}$&
     \textbf{65.4}   &   \textbf{58.2}    &   \textbf{68.5}    &   \textbf{49.4}    &   \textbf{65}  & \textbf{61.3} \\
     $^{5,\text{M}}$\FcNNfft & 
     \mean{16.77+   56.04+  32.29+  23.02+  29.48}&
     \mean{23.33+   18.85+  37.81+  11.35+  35.83}&
     \mean{33.44+   25.31+  38.44+  10.94+  36.46}&
     \mean{9.90+    25.31+  49.90+  32.71+  10.52}&
     \mean{58.13+   37.50+  60.83+  81.98+  34.17}&
     \mean{28.31+   32.60+  43.85+  32.00+  29.29}\\
     $^{5,\text{H}}$\FcNNfft & 
     \mean{22.29+   27.29+  36.56+  22.29+  23.96}&
     \mean{35.31+   13.33+  47.81+  18.13+  30.83}&
     \mean{12.19+   14.17+  36.04+  25.42+  45.73}&
     \mean{9.90+    32.92+  41.46+  52.60+  25.94}&
     \mean{71.56+   58.85+  83.96+  75.94+  38.65}&
     \mean{30.25+   29.31+  49.17+  38.88+  33.02}\\
\end{tabular}\label{tab:results}
}
    
\end{table}

\begin{table}[t!]
\caption{Confusion matrix of the best-performing network ($^{10}$CNN+FcNN), showing the label and prediction of all cross-validations with Hard Bone~(HB), Soft Bone~(SB), Fat~(F), Skin~(S), and Muscle~(M).}
\footnotesize
    \centering
    
    \begin{tabular}{r|| c c c c c | c}
             &   HB      & SB        &   F           &   S           &  M            & acc. \\
         \hline
         \hline
        HB  &   \sumbf{699+819+663+715+713}     &    \sumL{25+45+20+23+150}    &   \sumL{16+10+0+25+2}   &   \sumL{170+25+113+128+29}  & \sumL{0+11+114+19+16} & 79.3\%\\
        SB  &   \sumL{93+63+13+192+15} & \sumbf{601+817+767+673+895} & \sumL{1+10+81+32+0} & \sumL{134+20+2+5+0} & \sumL{81+0+47+8+0} & 82.5\% \\
        F   & \sumL{28+77+8+29+0}   & \sumL{42+43+1+0+1}    & \sumbf{675+567+858+697+741} & \sumL{72+90+20+59+69}   & \sumL{93+133+23+125+99} & 77.8\% \\
        S   &   \sumL{271+174+63+80+230} & \sumL{119+25+22+0+10} & \sumL{93+0+16+20+96}  & \sumbf{360+708+635+801+538}   & \sumL{67+3+174+9+36}  & 66.9\% \\
        M   & \sumL{3+0+2+13+106} & \sumL{35+58+8+4+29} & \sumL{64+177+228+62+112}  & \sumL{94+0+67+23+235} & \sumbf{714+675+605+808+428}   & 71\% \\
        \hline
         \multicolumn{6}{r |}{Mean acc.}  & \textbf{75.5\%} \\
    \end{tabular}

    \label{tab:confusion}
\end{table}

In general, using 10 consecutive acoustic waves performed better than using 1 or 5.
An exception was when only the frequency-dependent data as input were used. There, the $^5$\FcNNfft~ network, which used 5 consecutive acoustic waves, outperformed the $^1$\FcNNfft~ and $^{10}$\FcNNfft~ networks, which used the frequency spectrum of either 1 or 10 consecutive waves. 

The architecture with the worst performance was the ANN using the first three principal components of the frequency spectrum. It reached an accuracy of 30.4\% using $10$ consecutive waves.
The accuracy improved when relaxing the conditions. 
We created a deeper network with more neurons, added dropout layers, and used 323 principal components.
This network achieved an accuracy of 58\% using the spectrum of $10$ successive waves as input to the NN (\FcNNpca~, Figure~\ref{fig:model}).
Note that using the frequency spectrum instead of the principal components improved accuracy. 
Our optimized NN (\FcNNfft~, Figure~\ref{fig:model}), achieved an accuracy of 62.2\% using only the frequency spectrum of 5 consecutive waves. 
The accuracy improved when time-dependent data was used as an input to the CNN, reaching an accuracy of 70.8\% when using the \CNNtime~ with 10 consecutive measurements (see Table~\ref{tab:results}.
The best performing NN was CNN+FcNN (Figure~\ref{fig:model}), where we combined the time-dependent measurements and their Fourier transform, where with 10 consecutive waves, a classification accuracy of 75.5\%, as shown in Table~\ref{tab:results}. 
Furthermore, the confusion matrix of the best-performing model can be found in Table~\ref{tab:confusion}.

\subsection*{Grad-Cam (\CNNfft)}\label{sec:GraCam}
\begin{figure}[t]
    \centering
    \includegraphics[width=.475\textwidth]{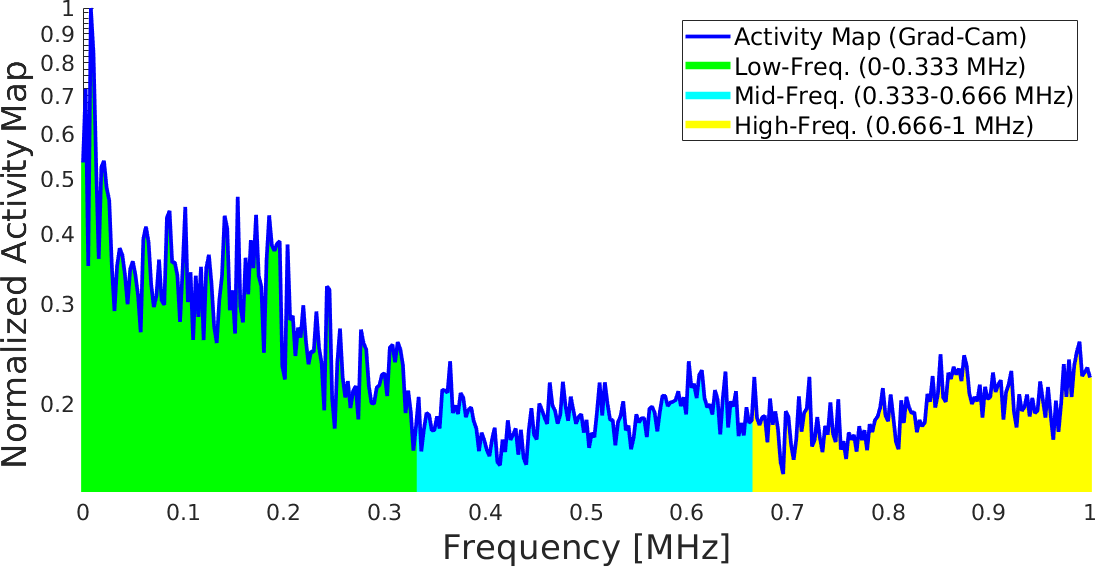}
    \caption{The figure visualizes the activation map resulting from \CNNfft. The higher the value, the more important the frequency is for the classification task. The green area under the blue curve is the largest (low-frequency), then the high-frequency area market as yellow, and the smallest area is the mid-frequency market with cyan. }
    \label{fig:activity}
\end{figure}

To analyze the effect of the frequency spectrum, we computed the activation map using the gradient of the convolutional layer of \CNNfft.
Note that only the frequency spectrum of 5 consecutive waves was considered because it was the best-performing network for the frequency domain in \FcNNfft.
To this end, a convolutional layer with a kernel size of one was added, resulting in the network \CNNfft.
We computed the activation map using only the correct prediction and show the resulting activation map in Figure~\ref{fig:activity}. 
The low-frequencies ($0-\SI{0.333}{\mega\Hz}$) had the most influence, as they had the largest area under the curve (green). 
This is confirmed by the fact that the low-frequency input to the NN outperforms the mid- and high-frequency inputs with an accuracy of 61.3\% ($^{5,\text{L}}$\FcNNfft).
Note that the yellow area in Figure~\ref{fig:activity} is slightly larger than the cyan area, which is also reflected in the accuracy in Table~\ref{tab:results}, where using the high-frequency spectrum ($^{5,\text{H}}$\FcNNfft) outperforms the mid-frequency spectrum ($^{5,\text{M}}$\FcNNfft) by almost 3\%.

\section{Conclusion}

Our NNs use the acoustic waves emitted during tissue ablation with a microsecond pulsed Er:YAG laser to classify the ablated tissue.
Our approach outperforms the method proposed in~\cite{herve2019}. 
We observe that using the first three principal components is insufficient to classify the tissues accurately. 
In addition, the accuracy is improved by using more consecutive waves and principal components.
Since the network using the frequency spectrum outperforms the network using the principal components as input, we conclude that the projected values of the principal components omit important information of the acoustic wave.
Another observation is that we can extract additional information from the acoustic wave by using the time-dependent wave and adding convolutional layers to the NN. 
The network performance is further improved by using both frequency- and time-dependent data as joint inputs.

Using the activation map (Grad-Cam), we find that the low frequencies are the most important for the network. 
We compared low-, mid-, and high-frequencies as input to the NN and confirmed that the low frequencies have the highest impact on the classification task. These results confirm the claims in~\cite{herve2019} that the low-frequencies are the most important ones. 
However, our transducer has a frequency range of $0.1-\SI{0.8}{\mega\Hz}$.
This indicates that the transducer we used for the experiment was not optimal.

\subsection*{Future Work}
We used only specimens consisting of isolated tissues (hard bone, soft bone, muscle, fat, and skin tissue). 
However, this setup is not feasible during surgery. 
Therefore, we plan to investigate the more challenging case where specimens consist of multiple tissue layers. 
In addition, the dependence of the water content of the tissue and its effect on the resulting acoustic wave during tissue ablation needs to be investigated, as well as the effect of different transducers and their position during the ablation process.

 \section*{Acknowledgment}
 We thank the Werner Siemens Foundation for funding the flagship project \mbox{MIRACLE}.
 We would like to thank Robin Sandk\"uhler, Florentin Bieder, Ferda Canbaz, Arsham Hamidi, and Sandra Drusová for helpful discussions.

\bibliographystyle{IEEEtran}
\bibliography{main.bib}

\EOD

\end{document}